\documentclass{jpsj-suppl}

\title{Nucleon Resonances Near 2 GeV}

\author{Jun He}

\inst{Theoretical Physics Division, Institute of Modern Physics, Chinese Academy of Sciences,Lanzhou~730000,China
}

\email{junhe@impcas.ac.cn}

\recdate{Sept XX, 2015}

\abst{The nucleon resonances near 2 GeV are investigated through the $\Sigma$(1385) and $\Lambda(1520)$ photoproductions within a Regge-plus-resonance approach based on the new experimental data released by the CLAS Collaboration. The $\Delta(2000)$ and the $N(2120)$ are found essential to reproduce the experimental data and should be assigned as second $[\Delta 5/2^+]$  and third $[N3/2^-]$ in the constituent quark model, respectively.  A calculation  of the binding energy and decay pattern supports that the $N(1875)$, which is listed in the PDG as the third $N3/2^-$ nucleon
resonance instead of the $N(2120)$, is from the $\Sigma(1385)K$ interaction rather than a three quark state.
 }

\kword{nucleon resonance, photoproduction, Regge-plus-resonance, CLAS experiment}

\begin{document}
\maketitle

\section{Introduction}

In the recent years, people's understanding of the nucleon resonances below 2 GeV has been deepened by the meson photoproduction experiments in facilities such as CLAS, MAMI, LEPS and GRAAL.  
However, due to the lack of precise experimental data the study about the nucleon resonances near 2 GeV, especially its internal structure, is scarce on both theoretical and experimental sides.  In the constituent
quark model (CQM), about two dozens of the nucleon resonances is predicted in this energy range. Only a few of them have been observed as shown in the Review of Particle Physics (PDG) with large uncertainties in
mass and width~\cite{Agashe:2014kda}.

Currently the single meson photoproductions are widely used to study the nucleon resonances. However, a high-mass nucleon resonance
is prone to decay into a meson with a baryon resonance, as well as  a ground baryon. Such decay should be observed in multi-meson productions, which provides a new way to detect the internal structure of high-mass nucleon resonance. Besides, the high thresholds of such processes will make the contribution from high-mass nucleon resonance more significant. 

In the literature, the radiative and strong decays of nucleon resonances to  $\Sigma$(1385), $\Lambda(1520)$ or $\Lambda(1405)$ have been studied in the CQM~\cite{Capstick:1998uh,Capstick:1992uc}.  The theoretically predicted large decay widths to $p\gamma$ and $K\Lambda(1520)$ inspire us to study the $N(2120)$ in the kaon photoproduction with the $\Lambda(1520)$ off proton target,
which has been confirmed by many theoretical analysis of the
$\Lambda(1520)$ photoproduction data~\cite{Xie:2010yk,He:2012ud}. Analogously, the $\Delta(2000)$ is expected to play an important role in the $\Sigma(1385)$ photoproduction.

Very recently, the high precision experimental data about
the kaon photoproduction with $\Sigma$(1385), $\Lambda(1520)$ and
$\Lambda(1405)$ were released by the CLAS collaboration~\cite{Moriya:2013hwg}, which
provides an opportunity to study high-mass nucleon resonances in these
processes.
In this work, 
the new CLAS data will be analyzed within a Regge-plus-resonance approach and
the roles played by nucleon resonances near 2 GeV will be studied.

\section{$\Sigma(1385)$ photoproduction and resonance $\Delta(2000)$}

In the CQM, a large amount of nucleon resonances in the
energy region considered in this work were predicted. In this work only those with large decay widths in $\gamma p$ and $\Sigma(1385)K$ channels will be considered, which are listed in Table \ref{Tab: 1385}. The results suggest that the $\Delta(2000)$ as second CQM state $[\Delta\frac{5}{2}^+]$ is essential to reproduce the experimental data. Without this resonance, the $\chi^2$ will increased by 4.5.

\begin{table*}[htp!]
\renewcommand\tabcolsep{0.515cm}
\renewcommand{\arraystretch}{1.}
\caption{The nucleon resonances considered in $\Sigma(1385)$ photoproduction. The mass $m_R$, helicity
	amplitudes $A_{1/2,3/2}$ and partial wave decay amplitudes
	$G(\ell)$  from Refs.~\cite{Capstick:1998uh,Capstick:1992uc} are in the unit of MeV, $10^{-3}/\sqrt{\rm{GeV}}$ and
	$\sqrt{\rm{MeV}}$, respectively. The last column is for the variation of $\chi^2$ after
turning off the corresponding nucleon resonance and refitting. In the full model $\chi^2=2.5$ per degree of freedom.}
\begin{tabular}{ll|rr|rr|rrr} \hline
 State & PDG&  $A^p_{1/2}$ &   $A^p_{3/2}$ &  $G(\ell_1)$ &
 $G(\ell_2)$  &$\delta\chi^2$\\\hline
 $[N\textstyle{3\over 2}^-]_3(1960)$ &$N(2120)$ & 36 &  $-43$  & $1.3 ^{+ 0.4}_{-
 0.4}$ & $1.4 ^{+ 1.3}_{- 1.3}$ &0.4\\
 $[N\textstyle{3\over 2}^-]_5(2095)$ & & $-9$   & $-14$  &
 $7.7 ^{+1.2}_{-1.2}$ &$-0.8$ $^{+0.7}_{-1.0}$    &0.1\\
  $[\Delta\textstyle{3\over 2}^-]_2(2080)$ & $\Delta(1940)$&
  $-20$  &$-6$  & $-4.1 ^{+ 4.0}_{- 1.5}$ &
  $-0.5 ^{+ 0.5}_{- 2.2}$   &0.0\\
$[\Delta\textstyle{3\over 2}^-]_3(2145)$ & & 0  &10 & 5.2 $\pm $ 0.4 &
$-1.9 ^{+ 1.2}_{- 4.0}$  &0.0\\
  $[\Delta\textstyle{5\over 2}^+]_2(1990)$ &$\Delta(2000)$  &
  $-10$ &$-28$ & 4.0 $^{+ 4.5}_{-
  4.0}$ & $-0.1 ^{+ 0.1}_{- 0.4}$   &4.5\\
\hline
\end{tabular}
\label{Tab: 1385}
\end{table*}

\begin{figure}[h!]
\begin{center}
  \includegraphics[ bb= 179 239 560 620 ,scale=0.58,clip]{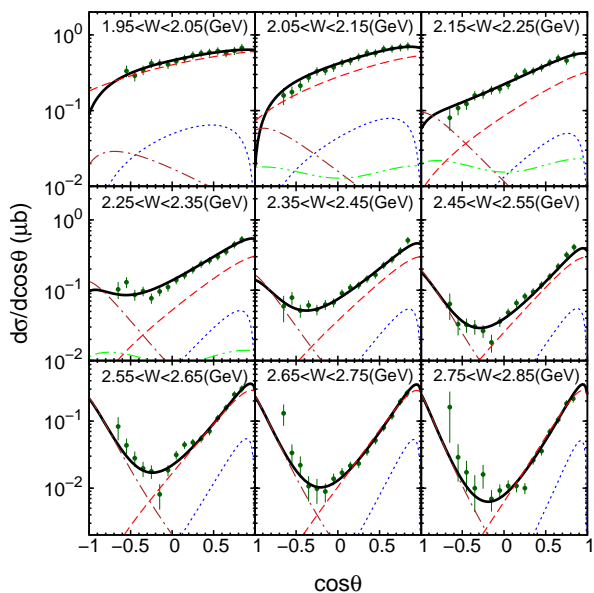}
  \end{center}
  \caption{The differential cross section $d\sigma/d\cos\theta$ for
the $\Sigma$(1385) photoproduction from proton as a function of
$\cos\theta$. The full (black), dashed (red),
dash-dotted (brown), dotted (blue) and dash-dot-dotted (green) lines are for the full model,
the contact term, the  $u$ channel, the $t$ channel and $\Delta(2000)$, respectively. The data are from
\cite{Moriya:2013hwg}.}
	  \label{Fig: dcs}
\end{figure}

The differential cross section of the $\Sigma$(1385)
photoproduction off proton is well reproduced as shown in
Fig.~\ref{Fig: dcs}. The contact term is dominant in the whole energy range considered. The Reggized $t$-channel  is important at forward angles while the $u$-channel is responsible to the behavior of differential cross section at backward angles.

\section{$\Lambda(1520)$ photoproduction and resonance $N(2120)$}

Analogous to previous section, only nucleon resonance with large decay widths to $\gamma p$ and $\Lambda(1520)K$ will be considered, which are listed in Table. \ref{Tab: 1385}. Among those nucleon resonates, the largest variation of $\chi^2$ is found after turning off the $N(2120)$, which is assigned as third nucleon resonance $[N\frac{3}{2}^-]$ in the CQM.

\begin{table*}[h!]
\renewcommand\tabcolsep{0.535cm}
\renewcommand{\arraystretch}{1.}
\caption{The nucleon resonances considered in the $\Lambda(1520)$ photoproduction.  $\chi^2=3.1$ in the full model. Notation as in Table~\ref{Tab: 1385}. }
\begin{tabular}{ll|rrrr|rr}
 \hline State  &PDG & $A^p_{1/2}$ &  $A^p_{3/2}$ &
  $G(\ell_1)$ &  $G(\ell_2)$   &  $\delta\chi^2$\\\hline
 $[N\textstyle{1\over 2}^-]_3(1945)$ & $N(1895)$ & 12  &  & 6.4 $^{+
 5.7}_{- 6.4}$ & & 0.0 \\
 $[N\textstyle{3\over 2}^-]_3(1960)$ &    $N(2120)$ & 36  & -43 & $-2.6 ^{+ 2.6}_{-
 2.8}$ & $-0.2 ^{+ 0.2}_{- 1.3}$ &0.5\\
 $[N\textstyle{5\over 2}^-]_2(2080)$ &$N(2060)$  & -3  & -14 & $-4.7 ^{+
 4.7}_{- 1.2}$ & $-0.3 ^{+ 0.3}_{- 0.8}$ &0.1  \\
  $[N\textstyle{5\over 2}^-]_3(2095)$ & & -2  & -6  & $-2.4 ^{+ 2.4}_{-
 2.0}$ & $-0.1 ^{+ 0.1}_{- 0.3}$ & 0.0\\
 $[N\textstyle{7\over 2}^-]_1(2090)$ &$N(2190)$ & -34  & 28  & $-0.5 ^{+ 0.4}_{- 0.6}$ & 0.0
 $^{+0.0}_{-0.0}$ & 0.1\\
 $[N\textstyle{7\over 2}^+]_2(2390)$ & & -14  & -11  & 3.1 $^{+ 0.8}_{- 1.2}$ & 0.3 $^{+
 0.3}_{- 0.2}$   &0.1\\
 \hline
\end{tabular}
\label{Tab: 1520}
\end{table*}

One can find that experimental data are well reproduced in our model at
all energies and angles. The dominant contributions
are from Born terms, among which the contact term plays the most important role.
The Reggized $t$ channel is important to
reproduce the behavior of the differential cross section at forward
angles especially at high energies. The $u$ channel is responsible to
the increase of the differential cross section with the decrease of
$\cos\theta$ at high energies.  However, the similar increase at low
energies is from $N(2120)$ contribution instead.

\begin{figure}[h!]
\begin{center}
  \includegraphics[ bb= 170 230 700 618 ,scale=0.53,clip]{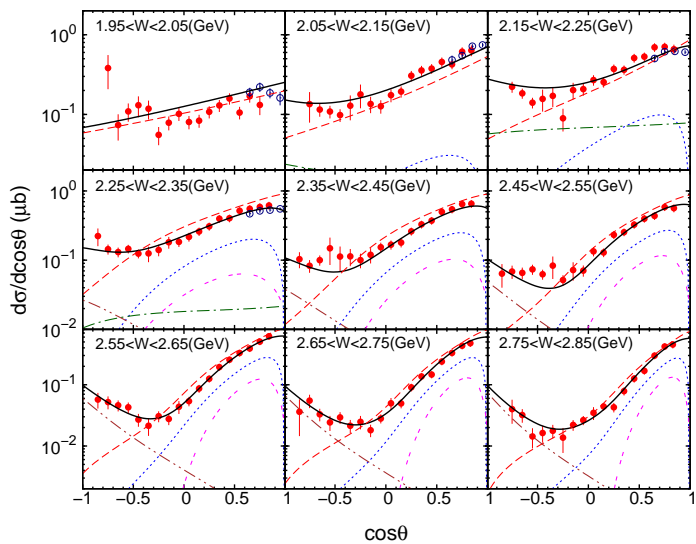}
  \caption{The differential cross section
	  $d\sigma/d\cos\theta$ with variation
	  of $\cos\theta$ for
the $\Lambda$(1520) photoproduction . The full (black), dashed (red), dotted
  (blue), dash-dot-dotted( brown), dash-dashed (magenta) and
  dash-dotted (darkgreen) lines are for full model, contact term, $K$ exchange $t$
  channel, $u$ channel, $K^*$ exchange $t$ channel and $N(2120)$
  . The full circle (red) and open circle (blue) are for
  CLAS13 data \cite{Moriya:2013hwg} and LEPS10 data
  \cite{Kohri:2009xe}.}
  \label{Fig: CLAS13}
\end{center}
\end{figure}

\section{Internal structure of the $N(2120)$}

The CLAS experimental data support the assignment of $N(2120)$ as the third nucleon resonance state with $J^P=3/2^-$ in the CQM so that there is no position to
settle the $N(1875)$, third $N3/2^-$ in the PDG. The closeness of the $N(1875)$ and the $\Sigma(1385) K$ threshold encourages us to interpret the $N(2120)$ as a $\Sigma(1385) K$ molecular state. With such assumption, the binding eerngy is calculated in a Bethe-Salpeter equation approach, and the wave function obtained is  used to study  the decay pattern of the $N(1875)$ through hadronic loop mechanism, which is listed in Table \ref{decay}.

\renewcommand\tabcolsep{0.4cm}
\renewcommand{\arraystretch}{1.}
\begin{table}[hbtp!]
\caption{The binding energies $E$ for $\Sigma^*K$ system with different cut off $\Lambda$
	The cut off
	$\Lambda$, binding energy and branch ratio are in the units of GeV,	MeV and \%, respectively.
\label{decay}}
\begin{center}
\begin{tabular}{c|rr|rrrrrr}\hline
$\Lambda$ & $E$ &  $\Gamma$ & $N\sigma$ &  $N\rho$ &
$N\omega$ &  $N\pi$ & $\Lambda K$ & $\Sigma K$\\\hline
  1.72   &   8  &   73&  55.8  &  4.7  &   14.0   &   22.6   &    2.3   &  0.6  \\
  1.80   &  28  &  155&  55.6  &  4.8  &   14.2   &   22.8   &    2.1   &  0.5  \\
  1.88   &  67  &  257&  54.9  &  5.1  &   14.9   &   22.9   &    1.8   &  0.4  \\\hline
  PDG \cite{Agashe:2014kda} &  $30^{+25}_{-25}$&        & $24^{+24}_{-24}$  &     $6^{+6}_{-6}$    &
  $20^{+4}_{-4}$  &   $7^{+6}_{-6}$   &    & $0.7^{+0.4}_{-0.4}$       \\
  BnGa \cite{Anisovich:2011fc}  &  $0^{+20}_{-20}$&    $200^{+20}_{-20}$    & $60^{+12}_{-12}$  &      &
  &   $3^{+2}_{-2}$   &   $4^{+2}_{-2}$ & $15^{+8}_{-8}$
  \\\hline
  CQM \cite{Capstick:1998uh}& -85 &   $324$  & & $57.1$  &   $12.3$   &
   $20.8$   &   $9.7$ & $0$       \\
\hline
\end{tabular}
\end{center}
\end{table}

The branch ratios of $N(1875)$ decay
are stable compared with binding energy. The $N\sigma$ channel is the most
important decay channel, about $55\%$, which is consistent with the
PDG suggested values $24\pm24\%$ and $60\pm12\%$ from the BnGa analysis.
The main decay channel of the $[N3/2^-]_3$ predicted in the CQM is
$N\rho$ which is much larger than other decay channels, which
conflict with the PDG values and the BnGa analysis.
Hence, the decay pattern of the $N(1875)$
supports the molecular state interpretation.

\section{Summary}

Based on an analysis of the new CLAS data of the $\Sigma(1385)$ and the $\Lambda(1520)$ photoproductions, the $\Delta(2000)$ and the $N(2120)$ are suggested to be assigned as second $[\Delta 5/2^+]$  and third $[N3/2^-]$ in the CQM, respectively. The $N(1875)$, which is listed in the PDG as the third $N3/2^-$ nucleon
resonance instead of the $N(2120)$, is from the $\Sigma(1385)K$ interaction rather than a three quark CQM state.

This project is  supported by 
the National Natural Science
Foundation of China (Grants No. 11275235).

\end{document}